\documentclass[aps,pra,twocolumn,superscriptaddress,longbibliography]{revtex4-1}

\usepackage{amsmath}
\usepackage{amssymb}
\usepackage{graphicx}
\usepackage[dvipsnames]{xcolor}
\usepackage[colorlinks,linkcolor=blue,citecolor=blue,urlcolor=blue]{hyperref}
\usepackage{stix}

\begin{document}

\renewcommand{\vec}[1]{\mathbf{#1}}
\newcommand{\gvec}[1]{\boldsymbol{#1}}
\newcommand{\iu}{\mathrm{i}}
\newcommand{\hc}{\hat{c}}
\newcommand{\hcd}{\hat{c}^\dagger}
\newcommand{\en}{\varepsilon}
\renewcommand{\pl}{\parallel}

%************************************************
%                    Start
%************************************************

\author{Michael Sch\"uler}
\email{schuelem@stanford.edu}
\affiliation{Stanford Institude for Materials and Energy Sciences (SIMES),
  SLAC National Accelerator Laboratory, Menlo Park, CA 94025, USA}
\author{Jacob A. Marks}
\affiliation{Stanford Institude for Materials and Energy Sciences (SIMES),
  SLAC National Accelerator Laboratory, Menlo Park, CA 94025, USA}
\affiliation{Physics Department, Stanford University, Stanford, CA 94035, USA}
\author{Yuta Murakami}
\affiliation{Department of Physics, Tokyo Institute of Technology, Meguro, Tokyo 152-8551, Japan}
\author{Chunjing Jia}
\affiliation{Stanford Institude for Materials and Energy Sciences (SIMES),
  SLAC National Accelerator Laboratory, Menlo Park, CA 94025, USA}
% \author{C. Das Pemmaraju}
% \affiliation{Stanford Institude for Materials and Energy Sciences (SIMES),
%   SLAC National Accelerator Laboratory, Menlo Park, CA 94025, USA}
% \author{Brian Moritz}
% \affiliation{Stanford Institude for Materials and Energy Sciences (SIMES),
%   SLAC National Accelerator Laboratory, Menlo Park, CA 94025, USA}
\author{Thomas P. Devereaux}
\affiliation{Stanford Institude for Materials and Energy Sciences (SIMES),
SLAC National Accelerator Laboratory, Menlo Park, CA 94025, USA}
\affiliation{Department of Materials Science and Engineering, Stanford University, Stanford, California 94305, USA}

\title{Gauge invariance of light-matter interactions in first-principle tight-binding models}

\begin{abstract}
  We study the different ways of introducing light-matter interaction in first-principle tight-binding (TB) models. The standard way of describing optical properties is the velocity gauge, defined by linear coupling to the vector potential. 
  In finite systems a transformation to represent the electromagnetic radiation by the electric field instead is possible, albeit subtleties arise in periodic systems. The resulting dipole gauge is a multi-orbital generalization of Peierl's substitution. In this work, we investigate accuracy of both pathways, with particular emphasis on gauge invariance, for TB models constructed from maximally localized Wannier functions. Focusing on paradigmatic two-dimensional materials, we construct first-principle models and calculate the response to electromagnetic fields in linear response and for strong excitations. Benchmarks against fully converged first-principle calculations allow for ascertaining the accuracy of the TB models. We find that the dipole gauge provides a more accurate description than the velocity gauge in all cases. The main deficiency of the velocity gauge is an imperfect cancellation of paramagnetic and diamagnetic current. Formulating a corresponding sum rule however provides a way to explicitly enforce this cancellation. This procedure corrects the TB models in the velocity gauge, yielding excellent agreement with dipole gauge and thus gauge invariance. 
\end{abstract}

\maketitle

\section{Introduction}

The impressive progress in tailoring ultrafast laser pulses has led to a surge of advanced spectroscopies on and control of condensed matter systems~\cite{basov_towards_2017}. Prominent examples of intriguing phenomena beyond linear response include nonlinear Bloch oscillations~\cite{schubert_sub-cycle_2014,reimann_subcycle_2018}, and photo-dressing the electronic structure in Floquet bands~\cite{wang_observation_2013,mahmood_selective_2016,de_giovannini_monitoring_2016,hubener_creating_2017,schuler_how_2020-1}. Another recent pathway to controlling the properties of materials is exploiting the quantum nature of the electromagnetic fields in cavities, thus creating novel light-matter systems~\cite{ruggenthaler_quantum-electrodynamical_2018,mazza_superradiant_2019}. 

Simulating the response of complex materials to (possibly strong) external fields proves challenging. Density functional theory (DFT) or time-dependent DFT (TDDFT) provides a path to treat materials including electronic correlations, although the accuracy is limited by the inevitable approximations to the exchange-correlation functional. Depending on the choice of the basis, incorporating electromagnetic fields via the minimal coupling $\hat{\vec{p}}\rightarrow \hat{\vec{p}} - q \vec{A}(\vec{r},t)$ ($\hat{\vec{p}}$ denotes the momentum operator, $\vec{A}(\vec{r},t)$ the vector potential) is straightforward. Upon converging with respect to the basis, this approach provides a first-principle route to optical properties \cite{pemmaraju_velocity-gauge_2018} and nonlinear phenomena~\cite{de_giovannini_monitoring_2016,tancogne-dejean_atomic-like_2018,tancogne-dejean_ultrafast_2018}.

However, there are many scenarios where a reduced set of bands is preferable, for instance when many-body techniques beyond DFT are employed. Typical examples are strongly correlated systems~\cite{golez_dynamics_2019,petocchi_hund_2019}, excitonic effects~\cite{attaccalite_real-time_2011,perfetto_pump-driven_2019}, or systems where scattering events (like electron-phonon) play a crucial role in the dynamics~\cite{sentef_examining_2013,molina-sanchez_temperature-dependent_2016,schuler_how_2020-1}.
The canonical way of introducing a small subspace is the tight-binding (TB) approximation. TB models are typically constructed by fitting a parameterization to a DFT calculation, or by constructing Wannier functions. While the former approach is straightforward, the thus obtained empirical TB models lack the information on the underlying orbitals and hence the light-matter coupling. In this context the Peierl's substitution~\cite{peierls_zur_1933,ismail-beigi_coupling_2001}, is often used to incorporate the external field. However, this approach neglects local inter-orbital transitions. Introducing matrix elements of the light-matter coupling in the minimal coupling scheme (velocity gauge) as fitting parameters is possible, but does not provide a way to construct them and generally breaks gauge invariance~\cite{foreman_consequences_2002}. 

In contrast, Wannierization of a subspace of the DFT electronic structure -- if possible -- provides a systematic way of constructing first-principle TB Hamiltonians including the orbital information. Expressing the Bloch states $|\psi_{\vec{k}\alpha}\rangle$ in terms of the Wannier functions allows to calculate the matrix elements of $\hat{\vec{p}}$ (velocity matrix elements) directly. However, typically the momentum operator is replaced in favor of the position operator~\cite{yates_spectral_2007} by employing the commutation relation
\begin{align}
    \label{eq:momentum_comm}
    \hat{\vec{p}} = \frac{m}{i\hbar} [\hat{\vec{r}},\hat{H}] \ ,
\end{align}
as the matrix elements of $\hat{\vec{r}}$ in the Wannier basis are directly obtained from the standard Wannierization procedure. Furthermore, Wannier models provide a straightforward way to express the Hamiltonian at any point in momentum space by Wannier interpolation, which greatly facilitates the otherwise costly calculation of optical transition matrix elements on dense grids. 

Treating the matrix elements of $\hat{\vec{r}}$ (dipole matrix elements) as the more fundamental quantity, it would be advantageous to express the Hamiltonian directly in terms of the dipoles instead of taking the detour via Eq.~\eqref{eq:momentum_comm}. In finite systems and within the dipole approximation (neglecting the spatial dependence of the field), this is achieved by the Power-Zienau-Woolley transformation to the dipole gauge, resulting in the light-matter interaction of form $\hat{H}_{\mathrm{LM}} = - q \vec{E}(t)\cdot \hat{\vec{r}}$. In periodic systems, the operator $\hat{\vec{r}}$ is ill-defined in the Bloch basis, but a multi-center generalization of the Power-Zienau-Woolley transformation can be constructed~\cite{golez_multiband_2019,li_electromagnetic_2020,mahon_microscopic_2019} as detailed below. Working within a localized Wannier basis also provides a natural way to capture the magnetoelectric response of solids~\cite{mahon_magnetoelectric_2020,mahon_magnetoelectric_2020-1}. 

In principle, all of the mentioned schemes for incorporating light-matter interaction are equivalent and thus gauge invariant. In practice however, breaking the completeness of the band space by truncation introduces artifacts and a dependence on gauge. In this work, we compare the schemes of introducing light-matter coupling to TB models -- (i) in the dipole gauge (TB-DG), and (ii) in the velocity gauge (TB-VG). In particular we focus on the current as a fundamental observable determining the optical properties. We study the optical conductivity within the linear response formalism and, furthermore, the resonant excitations beyond linear response. All results are benchmarked against accurate first-principle calculations in either the plane-wave or the real-space representation of the Bloch wave-functions (which does not invoke any approximation with respect to the basis if converged with respect to the grid spacing).

This paper is organized as follows. In Sec.~\ref{sec:lm} we introduce the light-matter interaction in the different gauges. Starting from the velocity gauge (Sec.~\ref{subsec:lm_velo}) we work out the transformation to the dipole gauge for completeness (Sec.~\ref{subsec:lm_len}), with particular emphasis on the gauge invariance. In Sec.~\ref{sec:examples} we systematically investigate the accuracy of gauges when applied to first-principle TB models. 
We restrict our focus to typical two-dimensional systems, and calculate the optical conductivity (Sec.~\ref{subsec:optcond}) and Berry curvature (Sec.~\ref{subsec:berry}). Finally, we study nonlinear excitations (Sec.~\ref{sec:nonlin}). We use atomic units (a.u.) throughout the paper unless stated otherwise.

\section{Light-matter interaction in periodic systems \label{sec:lm}}

\subsection{Light-matter interaction in the velocity gauge \label{subsec:lm_velo}}

Here we recapitulate the form of the light-matter interacting arising from the minimal coupling principle.
Let us consider a crystalline solid with the periodic (single-particle) potential $v(\vec{r})$, which we take to be the Kohn-Sham potential obtained from DFT in the examples below. The Hamiltonian $\hat{h} = \frac{\hat{\vec{p}}^2}{2} + v(\vec{r})$ defines the eigenstates $\hat{h}| \psi_{\vec{k}\alpha} \rangle = \en_{\alpha}(\vec{k})| \psi_{\vec{k}\alpha} \rangle$. By virtue of the Bloch theorem, the periodic part $u_{\vec{k}\alpha}(\vec{r})$ is defined by $\psi_{\vec{k}\alpha}(\vec{r}) = e^{i \vec{k}\cdot \vec{r}} u_{\vec{k}\alpha}(\vec{r})$. Introducing the Bloch Hamiltonian $\hat{h}(\vec{k}) = e^{-i \vec{k}\cdot \vec{r}} \hat{h} e^{i \vec{k}\cdot \vec{r}}$ the periodic functions are obtained from $\hat{h}(\vec{k})|u_{\vec{k}\alpha} \rangle = \en_{\alpha}(\vec{k})|u_{\vec{k}\alpha} \rangle$. 

An electromagnetic wave interacting with the electrons in the sample can be represented by the vector potential $\vec{A}(t)$, which we assume to be spatially homogeneous. This is known as the dipole approximation, which holds as long as the wave length of the light is significantly larger than the extent of a unit cell. The minimal coupling $\hat{\vec{p}}\rightarrow \hat{\vec{p}} - q \vec{A}(t)$ ($q=-e$ is the charge of an electron) gives rise to the time-dependent Hamiltonian 
\begin{align}
  \label{eq:ham_mincoup}
  \hat{h}(t) = \frac12 (\hat{\vec{p}}-q \vec{A}(t))^2 + v(\vec{r}) \ .
\end{align}
If the Bloch wave-functions $\psi_{\vec{k}\alpha}(\vec{r})$ for (partially) occupied bands $\alpha$ are known, the time-dependent wave-functions can directly be obtained from the time-dependent Schr\"odinger equation (TDSE) $ i \partial_t \phi_{\vec{k}\alpha}(\vec{r},t) = \hat{h}(t)\phi_{\vec{k}\alpha}(\vec{r},t)$ with $\phi_{\vec{k}\alpha}(\vec{r},t=0) = \psi_{\vec{k}\alpha}(\vec{r})$. The averaged electronic current is calculated from the kinematic momentum operator $\hat{\vec{p}}_\mathrm{kin} = \hat{\vec{p}} - q \vec{A}(t)$:
\begin{align}
	\label{eq:current_rsp}
	\vec{J}(t) = \frac{1}{N}\sum_{\vec{k}} f_{\alpha}(\vec{k}) \langle \phi_{\vec{k}\alpha}(t)| \hat{\vec{p}} - q \vec{A}(t) | \phi_{\vec{k}\alpha}(t) \rangle \ ,
\end{align}
where $f_\alpha(\vec{k})$ denotes the occupation of the corresponding Bloch state; $N$ is the number of momentum points (or supercells, equivalently). In absence of spin-orbit coupling (SOC), Eq.~\eqref{eq:current_rsp} represents the current per spin, while $| \phi_{\vec{k}\alpha}(t) \rangle$ should be understood as a spinor in the case of SOC. 

Provided the time-dependent Bloch wave-functions are represented on a dense enough grid and the TDSE is solved with sufficient accuracy, the current~\eqref{eq:current_rsp} is the exact (independent particle) current. Let us now introduce a finite reduced band basis. All operators are expressed in the  basis of the corresponding Bloch states $|\psi_{\vec{k}\alpha}\rangle$.
The matrix elements of the time-dependent Hamiltonian~\eqref{eq:ham_mincoup} $h_{\alpha \alpha^\prime}(\vec{k},t) = \langle \psi_{\vec{k}\alpha} | \hat{h}(t) | \psi_{\vec{k}\alpha^\prime} \rangle$ are given by
\begin{align}
    \label{eq:sp_ham_velo}
    h_{\alpha\alpha^\prime}(\vec{k},t) = \en_\alpha(\vec{k})\delta_{\alpha \alpha^\prime}
    -q \vec{A}(t)\cdot\vec{v}_{\alpha\alpha^\prime}(\vec{k}) + \frac{q^2}{2} \vec{A}(t)^2\delta_{\alpha \alpha^\prime} \ .
\end{align}
Here, the last term denotes the diamagnetic coupling, which reduces to a pure phase factor in the dipole approximation.
In Eq.~\eqref{eq:sp_ham_velo} we have introduced the velocity matrix elements 
\begin{align}
    \label{eq:velo_elemk}
    \vec{v}_{\alpha\alpha^\prime}(\vec{k}) &= \langle \psi_{\vec{k}\alpha} | \hat{\vec{p}} | \psi_{\vec{k}\alpha^\prime} \rangle = -i \langle \psi_{\vec{k}\alpha} | [\hat{\vec{r}},\hat{h}] | \psi_{\vec{k}\alpha^\prime} \rangle \nonumber \\
    &=  \langle u_{\vec{k}\alpha} | \nabla_{\vec{k}} \hat{h}(\vec{k}) | u_{\vec{k}\alpha^\prime}\rangle  \ .
 \end{align}
Although a direct calculation of the velocity matrix elements~\eqref{eq:velo_elemk} is possible, in practical calculations (especially in the context of first-principle treatment) it is convenient to split into intra- and interband contributions.
One can show~\cite{yates_spectral_2007} that Eq.~\eqref{eq:velo_elemk} is equivalent to
\begin{align}
    \label{eq:velo_elemk_2}
    \vec{v}_{\alpha\alpha^\prime}(\vec{k}) = \nabla_{\vec{k}} \en_\alpha(\vec{k})\delta_{\alpha \alpha^\prime} - i \left(\en_{\alpha^\prime}(\vec{k}) - \en_{\alpha}(\vec{k})\right) \vec{A}_{\alpha\alpha^\prime}(\vec{k}) \ .
\end{align}
Here, $\vec{A}_{\alpha\alpha^\prime}(\vec{k}) = i \langle u_{\vec{k}\alpha} | \nabla_{\vec{k}}u_{\vec{k}\alpha^\prime} \rangle$ denotes the Berry connection.
Note that the equivalence of Eq.~\eqref{eq:velo_elemk} and Eq.~\eqref{eq:velo_elemk_2} is, strictly speaking, an approximation assuming
a complete set of Bloch states.
In the Bloch (band) basis, the total current is obtained by combining the paramagnetic and diamagnetic current:
\begin{align}
	\label{eq:total_current_velo}
	\vec{J}^\mathrm{VG}(t)&= \frac{q}{N}\sum_{\vec{k}}\sum_{\alpha \alpha^\prime} \left( \vec{v}_{\alpha\alpha^\prime}(\vec{k}) - q \vec{A}(t) \delta_{\alpha \alpha^\prime}\right) \rho_{\alpha^\prime\alpha}(\vec{k},t) \\ 
    &\equiv \vec{J}^\mathrm{p}(t) + \vec{J}^\mathrm{dia}(t) \nonumber
    \ .
\end{align}
Here, $\rho_{\alpha \alpha^\prime}(\vec{k},t)$ denotes the single-particle density matrix (SPDM), which is defined by the initial condition $\rho_{\alpha\alpha^\prime}(\vec{k},t=0) = f_{\alpha}(\vec{k})\delta_{\alpha \alpha^\prime}$ and the standard equation of motion.

% The time evolution is determined by the TDSE in the respective Bloch basis. For later convenience we formulate the dynamics in terms of the single-particle density matrix (SPDM) $\rho_{\alpha \alpha^\prime}(\vec{k},t)$, which obeys
% \begin{align}
% 	\label{eq:eom_spdm_velo}
% 	\frac{d}{d t} \gvec{\rho}(\vec{k},t) = -i \left[\vec{h}(\vec{k},t),\gvec{\rho}(\vec{k},t) \right]  
% \end{align}
% with $\rho_{\alpha\alpha^\prime}(\vec{k},t=0) = f_{\alpha}(\vec{k})\delta_{\alpha \alpha^\prime}$. Here, bold-face symbols denote compact matrix notation. From the time-depedent SPDM one can calculate the total electronic current by
% \begin{align}
% 	\label{eq:total_current_velo}
% 	\vec{J}(t)&= \frac{q}{N}\sum_{\vec{k}}\sum_{\alpha \alpha^\prime} \left( \vec{v}_{\alpha\alpha^\prime}(\vec{k}) - q \vec{A}(t) \delta_{\alpha \alpha^\prime}\right) \rho_{\alpha^\prime\alpha}(\vec{k},t) \\ 
%     &\equiv \vec{J}^\mathrm{p}(t) + \vec{J}^\mathrm{dia}(t) \nonumber
%     \ .
% \end{align}
% We remark that the first term (second) in Eq.~\eqref{eq:total_current_velo} is referred to as paramagnetic (diamagnetic) current; only together, one obtains a gauge-invariant current satisfying the equation of continuity~\cite{stefanucci_nonequilibrium_2013}.

\subsubsection{Wannier representation}

Calculating the Berry connections $\vec{A}_{\alpha\alpha^\prime}(\vec{k})$ is numerically challenging, as derivatives with respect to $\vec{k}$ are often ill-defined on a coarse grid of the Brillouin zone. This problem can be circumvented by switching to the Wannier representation
\begin{align}
    \label{eq:bloch_basis}
     | \psi_{\vec{k}\alpha} \rangle = \frac{1}{\sqrt{N}}\sum_{\vec{R}}e^{i \vec{k}\cdot\vec{R}}\sum_{m}C_{m\alpha}(\vec{k})
     |m\vec{R}\rangle \ ,
\end{align}
where $w_m(\vec{r}-\vec{R})=\langle \vec{r}|m\vec{R}\rangle$ denote the Wannier functions (WFs). At this point we invoke an important assumption: the WFs are assumed to be sufficiently localized, such that $\int d \vec{r}\, |\vec{r} w_m(\vec{r})|^2$ remains finite. As detailed in ref.~\onlinecite{yates_spectral_2007}, the Berry connection can then be expressed as
\begin{align}
    \label{eq:berryconect_2}
    \vec{A}_{\alpha\alpha^\prime}(\vec{k}) = \sum_{m m^\prime} C^*_{m\alpha}(\vec{k}) \left[
    \vec{D}_{m m^\prime}(\vec{k}) + i \nabla_{\vec{k}}\right]C_{m^\prime\alpha^\prime}(\vec{k}) \ .
\end{align}
The derivative in Eq.~\eqref{eq:berryconect_2} can then be replaced by an equivalent sum-over-states expression~\cite{yates_spectral_2007}.
Here we have defined the Fourier-transformed dipole operator
\begin{align}
    \label{eq:dipole_op}
    \vec{D}_{m m^\prime}(\vec{k}) = \sum_{\vec{R}} e^{i\vec{k}\cdot \vec{R}} \vec{D}_{m 0 m^\prime \vec{R}} \ ,
\end{align}
where $\vec{D}_{m \vec{R} m^\prime \vec{R}^\prime} = \langle m \vec{R} | \vec{r} - \vec{R} | m^\prime \vec{R}^\prime\rangle$ define the cell-centered dipole matrix elements. Note that they are well defined for sufficiently localized WFs. 

Eq.~\eqref{eq:total_current_velo} is independent of the choice of the band basis; hence, one can replace the Bloch bands by the basis spanned by the Wannier orbitals by replacing $\alpha\rightarrow m$. Note that the velocity matrix elements~\eqref{eq:velo_elemk_2} transform according to $\vec{v}_{m m^\prime}(\vec{k})=\sum_{\alpha \alpha^\prime} C_{m \alpha}(\vec{k}) \vec{v}_{\alpha\alpha^\prime}(\vec{k}) C^*_{m^\prime \alpha^\prime}(\vec{k})$, while the intraband current and the Berry connection term individually can not be transformed by a unitary transformation due to the derivative in momentum space. Without loss of generality, we assume the WFs to be orthogonal.

\subsubsection{Static limit of the current response}

Both the paramagnetic and the diamagnetic current contribute to the gauge-invariant total current. For an insulator, the total current in the linear-response regime in the direct-current (DC) limit must vanish in the zero-temperature limit, which amounts to paramagnetic and diamagnetic contributions canceling out. This defines an important sum rule for the velocity matrix elements~\eqref{eq:velo_elemk_2}. 
Let us consider the paramagnetic current-current response function 
\begin{align}
    \label{eq:para_resp}
    \chi^\mathrm{p}_{\mu \nu}(t) = -i \langle \left[\hat{J}^\mathrm{p}_\mu(t), \hat{J}^\mathrm{p}_\nu(0) \right] \rangle \ , 
\end{align}
where the operators on the right-hand side are understood in the Heisenberg picture ($\mu,\nu=x,y,z$ are the Cartesian directions.).  The response function~\eqref{eq:para_resp} defines the paramagnetic current by
\begin{align}
    J^\mathrm{p}_\mu(t) = \sum_{\nu}\int^t_{-\infty}\! d t^\prime\, \chi^\mathrm{p}_{\mu \nu}(t-t^\prime) A_\nu(t^\prime) \ ,
\end{align}
while the diamagnetic current becomes $J^\mathrm{dia}_\mu(t) = -n q^2 A_\mu(t)$ in linear response ($n$ is the number of particles per unit cell). Fourier transforming and requiring for total current $J_\mu(\omega=0)=0$ yields the sum rule
\begin{align}
    \label{eq:para_resp_fsum}
    \sum_\mu \chi^\mathrm{p}_{\mu\mu}(\omega=0)  = - n q^2 \ .
\end{align}
The sum rule~\eqref{eq:para_resp_fsum} holds for the fully interacting system. For noninteracting electrons Eq.~\eqref{eq:para_resp_fsum} reduces to
\begin{align}
    \label{eq:para_resp_free_fsum}
    f\equiv\frac{2}{N}\sum_{\vec{k}} \sum_{\alpha\ne\alpha^\prime} f_{\alpha}(\vec{k}) (1-f_{\alpha^\prime}(\vec{k}) ) 
    \frac{|\vec{v}_{\alpha\alpha^\prime}(\vec{k})|^2}{\en_{\alpha^\prime}(\vec{k}) - \en_{\alpha}(\vec{k})} = n  \ .
\end{align}
The relation Eq.~\eqref{eq:para_resp_free_fsum} provides an important criterion for the velocity matrix elements for assessing the completeness of the band space. 
Furthermore, the violation of the sum rule~\eqref{eq:para_resp_free_fsum} and thus of Eq.~\eqref{eq:para_resp_fsum} gives rise to spurious behavior of the optical conductivity, which is obtained from
\begin{align}
	\label{eq:optcond}
	\sigma_{\mu\nu}(\omega) = \frac{1}{i \omega} \left(\chi^\mathrm{p}_{\mu\nu}(\omega) - n q^2 \delta_{\mu\nu} \right) \ .
\end{align}
In particular, $\mathrm{Im}[\sigma_{\mu\nu}(\omega)] \propto 1/\omega$ for $\omega \rightarrow 0$ if $f\ne n$.
In general, sum-of-states expressions such as Eq.~\eqref{eq:para_resp_free_fsum} are slowly converging with respect to the number of bands included. 
Below we will exemplify this behavior and discuss how to cure this artifact of an (inevitably) incomplete Bloch basis.

\subsection{Light-matter interaction in the dipole gauge\label{subsec:lm_len}}

In finite systems, the dipole gauge is obtained by a unitary transformation of the type $\hat{U}(t) = \exp[-i q \vec{A}(t)\cdot \vec{r}]$. Applying this time-dependent transformation to the Hamiltonian~\eqref{eq:ham_mincoup}, we obtain
\begin{align}
	\label{eq:len_finite}
	\hat{h}_\mathrm{LG}(t) &= \hat{U}(t)\hat{h}(t) \hat{U}^\dagger(t) + (i \partial_t \hat{U}(t))\hat{U}^\dagger(t) \nonumber \\
	&= \frac{\hat{\vec{p}}^2}{2} + v(\vec{r}) - q \vec{E}(t)\cdot \vec{r} \ ,
\end{align}
where $\vec{E}(t)= - \dot{\vec{A}}(t)$ denotes the electric field. The extension to periodic systems and the corresponding Bloch states requires a few modifications. There is one subtle point which has to be taken care of: the dipole operator $\vec{r}$ (and any spatial operator without cell periodicity) is ill-defined with respect to the Bloch Basis. However, the dipole operator with respect to WFs ($\vec{D}_{m \vec{R} m^\prime \vec{R}^\prime}$) -- which defines the Berry connection via Eq.~\eqref{eq:dipole_op} and Eq.~\eqref{eq:berryconect_2} -- is well defined due to the localized nature of the WFs. Thus, $\vec{D}_{m \vec{R} m^\prime \vec{R}^\prime}$ and the Hamiltonian in Wannier representation $T_{m\vec{R}n\vec{R}^\prime} = \langle m \vec{R} | \hat{h} |n\vec{R}^\prime\rangle$ will be the constituents of the dipole gauge formulation.

\begin{figure*}[t]
\center
\includegraphics[width=\textwidth]{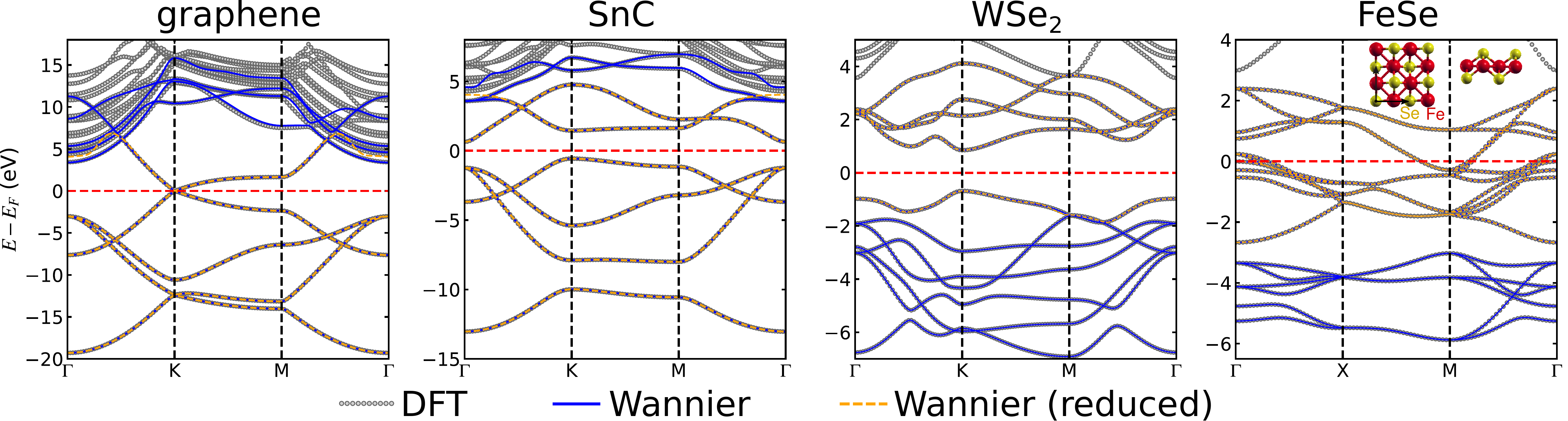}
\caption{Calculated band structures along typical paths in the respective Brillouin zone for the four considered systems. The energy scale is chosen relative to the Fermi energy $E_F$ (red dashed line). The inset for FeSe illustrates the geometry and chosen unit cell.  \label{fig:bands_wan}}
\end{figure*}

\subsubsection{Transformation to the dipole gauge}

Based on the dipole operator in Wannier representation we can define a similar unitary transformation as above. In the Wannier basis, we define
\begin{align}
\label{eq:unitary_len}
U_{m\vec{R}m^\prime\vec{R}^\prime}(t) = \langle m \vec{R} | e^{-\iu q \vec{A}(t)\cdot(\vec{r}-\vec{R})} | m^\prime\vec{R}^\prime\rangle \ .
\end{align}
Note that for Eq.~\eqref{eq:unitary_len} to be unitary, we assume $\vec{D}_{m\vec{R}m^\prime\vec{R}^\prime} = \vec{D}^*_{m^\prime\vec{R}^\prime m\vec{R}}$.

Transforming the time-dependent Hamiltonian using the transformation~\eqref{eq:unitary_len} yields 
\begin{align}
\label{eq:ham_sp_dipole_new}
\widetilde{h}_{m\vec{R} m^\prime\vec{R}^\prime}(t) = e^{i q \vec{A}(t)\cdot (\vec{R}-\vec{R}^\prime)} \left[ T_{m\vec{R} m^\prime\vec{R}^\prime} - q \vec{E}(t)\cdot \vec{D}_{m\vec{R}m^\prime\vec{R}^\prime}\right] \ .
\end{align}
Details are presented in Appendix~\ref{app:trans_len}. The additional phase factor in front of the field-free Wannier Hamiltonian $T_{m\vec{R}m^\prime \vec{R}^\prime}$ is the usual Peierl's phase factor~\cite{peierls_zur_1933}. Fourier transforming to momentum space, we obtain
\begin{align}
    \label{eq:spham_trans}
	\widetilde{h}_{m m^\prime}(\vec{k},t) = T_{m m^\prime}(\vec{k}-q\vec{A}(t)) - q \vec{E}(t)\cdot \vec{D}_{m m^\prime}(\vec{k}- q \vec{A}(t)) \ .
\end{align}
Here, $T_{m m^\prime}(\vec{k})$ is the Fourier-transformed Hamiltonian $T_{m m^\prime}(\vec{k})=\sum_{\vec{R}} e^{i \vec{k}\cdot\vec{R}} \langle m 0 | \hat{h}|m^\prime \vec{R}\rangle$. Eq.~\eqref{eq:spham_trans} can be understood as generalization of the Peierl's substitution for multiband systems. The density matrix in dipole gauge obeys the equation of motion according to the Hamiltonian~\eqref{eq:spham_trans} with the initial condition $\widetilde{\rho}_{m m^\prime}(\vec{k},t=0)=\sum_{\alpha} C_{m\alpha}(\vec{k})f_\alpha(\vec{k}) C^*_{m^\prime \alpha}(\vec{k})$. For vanishing field $\vec{A}(t)$ the density matrix in the different gauges is identical: $\widetilde{\rho}_{m m^\prime}(\vec{k},t) = \rho_{m m^\prime}(\vec{k},t)$. For $\vec{A}(t)\ne 0$ this equivalence is broken. In particular, the orbital occupation differes $\rho_{mm}(\vec{k},t) \ne \widetilde{\rho}_{mm}(\vec{k},t)$. This also leads to difference in the band occupation when transforming into the band basis. This gauge dependence of the density matrix does not affect any observables.

Note that any additional spatial operators entering the Hamiltonian are invariant by this unitary transformation. In particular, the Coulomb interaction is unaffected, which can be shown by carrying out the analogous steps on the level of the many-body Hamiltonian.

\subsubsection{Total current in the dipole gauge\label{subsubsec:len_curr}}

The expression for the current in the dipole gauge can be derived from the minimal coupling formulation~\eqref{eq:total_current_velo}. As for the Hamiltonian, the strategy is to express the momentum operator as $\hat{\vec{p}} = -i [\vec{r}, \hat{h}(t)]$ and express the position operator in the Wannier representation, $\vec{r}\rightarrow \vec{D}_{m\vec{R}m^\prime \vec{R}^\prime}$. The derivation is presented in Appendix~\ref{subsec:len_curr}. One obtains
\begin{align}
	\label{eq:curr_dip_wan}
	\vec{J}^\mathrm{LG}(t) = \vec{J}^\mathrm{disp}(t) + \vec{J}^\mathrm{dip}(t) \ ,
\end{align}
where 
\begin{align}
	\label{eq:curr_dip_wan_disp}
	\vec{J}^\mathrm{disp}(t) = \frac{q}{N}\sum_{\vec{k}} \sum_{m m^\prime} \nabla_{\vec{k}} 
	\widetilde{h}_{m m^\prime}(\vec{k},t) \widetilde{\rho}_{m^\prime m}(\vec{k},t) 
\end{align}
is the contribution related to the dispersion of the time-dependent Hamiltonian~\eqref{eq:spham_trans}. The second contribution arises from temporal variation of the polarization
\begin{align}
	\label{eq:pol_wan}
	\vec{P}(t) = \frac{q}{N}\sum_{\vec{k}} \sum_{m m^\prime} D_{m m^\prime}(\vec{k}-q\vec{A}(t)) 
	\widetilde{\rho}_{m^\prime m}(\vec{k},t) \ ,
\end{align}
by $\vec{J}^\mathrm{dip}(t) = d \vec{P}(t) / d t$. Under the assumptions stated above, gauge-invariance is guaranteed, i.\,e. $\vec{J}^\mathrm{VG}(t) = \vec{J}^\mathrm{LG}(t)$. For an incomplete set of WFs, the equivalence of Eq.~\eqref{eq:curr_dip_wan} and \eqref{eq:total_current_velo} are only approximate. In contrast to the velocity gauge, the cancellation of paramagnetic and diamagnetic current (which can not be separated in the dipole gauge) for an insulator at zero temperature is built in. Indeed, it can be shown (see Appendix~\ref{subsec:len_static}) that $\vec{J}^\mathrm{LG}(\omega = 0) = 0$ in linear response to a DC field is fulfilled by construction.

\section{First principle examples\label{sec:examples}}

In principle, the current within the velocity gauge~\eqref{eq:total_current_velo} and the dipole gauge~\eqref{eq:curr_dip_wan} is identical. In practice, truncating the number of bands introduces artifacts, which result in differences between the gauges and deviations from the exact dynamics. \emph{A priori} it is not clear which gauge is more accurate upon reducing the number of bands. Hence, we investigate the performance of both the dipole gauge and the velocity gauge in context of TB Hamiltonians, which are derived from first-principle calculations. This route also allows for comparing to converged first-principle treatment as a benchmark. 

For simplicity, we focus on a range of two-dimensional (2D) materials, albeit there is no inherent restriction. We start from graphene as the paradigm example of 2D systems and a Dirac semimetal. Substituting one carbon atom per unit cell breaks inversion symmetry and opens a gap~\cite{novoselov_two-dimensional_2005,geim_rise_2007}, making the system a (topologically trivial) insulator. As another example, we study SnC, which is thermally stable as a monolayer~\cite{hoat_structural_2019}. This material is also in the spotlight for the possibility to engineer the gap by strain~\cite{lu_tuning_2012}. We also consider monolayer WSe$_2$ as a prominent example of transition metal dichalcogenides (TMDCs). Finally, we study a monolayer of FeSe as a representative of a non-hexagonal structure. While free-standing FeSe is not stable, the layered structure renders a monolayer a good approximation to thin films, which are a prominent example of a high-temperature superconductor~\cite{lee_interfacial_2014,guterding_basic_2017,sentef_cavity_2018-1}. 

We performed first-principle DFT calculations based on the local-density approximation (LDA) using the {\sc Quantum espresso} code~\cite{giannozzi_quantum_2009}, and separately with the \textsc{Octopus} code~\cite{andrade_real-space_2015,tancogne-dejean_octopus_2020}. The consistency of the results has been checked. We used optimized norm-conserving pseudopotentials from the \textsc{PseudoDojo} project~\cite{van_setten_pseudodojo_2018}. In all cases, the self-consistent DFT calculation was performed with a $12\times 12$ Monkhorst-Pack sampling of the Brillouin zone. For the calculations with {\sc Quantum espresso} we used a supercell of 50 a.u. in the perpendicular direction, ensuring convergence of the relevant bands. Similarly, the \textsc{Octopus} calculations were performed with periodic boundary conditions in the plane, while the 50 a.u. long simulation box with open boundary conditions in perpendicular direction is chosen.

For constructing a first-principle TB model, we used the {\sc Wannier90} code~\cite{mostofi_updated_2014} to obtain maximally localized WFs (MLWFs) and a corresponding Wannier Hamiltonian for each system. For graphene, we include the $sp^2$, $p_z$ and a subset of $d$ orbitals, which allows to well approximate 9 bands (see Fig.~\ref{fig:bands_wan}). The analogous set of orbitals is chosen for SnC. A reduced model can be obtained by omitting the $d$ orbitals. For WSe$_2$ we included the W-$d$ orbitals and the Se-$p$ orbital; excluding the latter orbitals defines the reduced model. Similarly, the TB model for FeSe is constructed by choosing $d$ orbitals on Fe and $p$ orbital on Se sites. For clarity we focus on the extended TB models; results for the reduced models are shown in Appendix~\ref{app:reduced}. Fig.~\ref{fig:bands_wan} compares the first-principle band structure to the thus obtained TB models. \ 

We study optical properties and nonlinear dynamics. As we focus on the light-matter interaction itself, we treat the electrons as independent particles at this stage, thus excluding excitonic features. We also exclude any SOC. The dipole matrix elements $\vec{D}_{m\vec{R}m^\prime\vec{R}^\prime}$ are directly obtained from the output of {\sc Wannier90}. For calculating the velocity matrix elements according to Eq.~\eqref{eq:velo_elemk_2}, we extracted the calculation of the Berry connection~\eqref{eq:berryconect_2} from internal subroutines of {\sc Wannier90} into a custom code, taking $\vec{D}_{m\vec{R}m^\prime\vec{R}^\prime}$ and the Wannier Hamiltonian as input.

\begin{figure*}[t]
\center
\includegraphics[width=\textwidth]{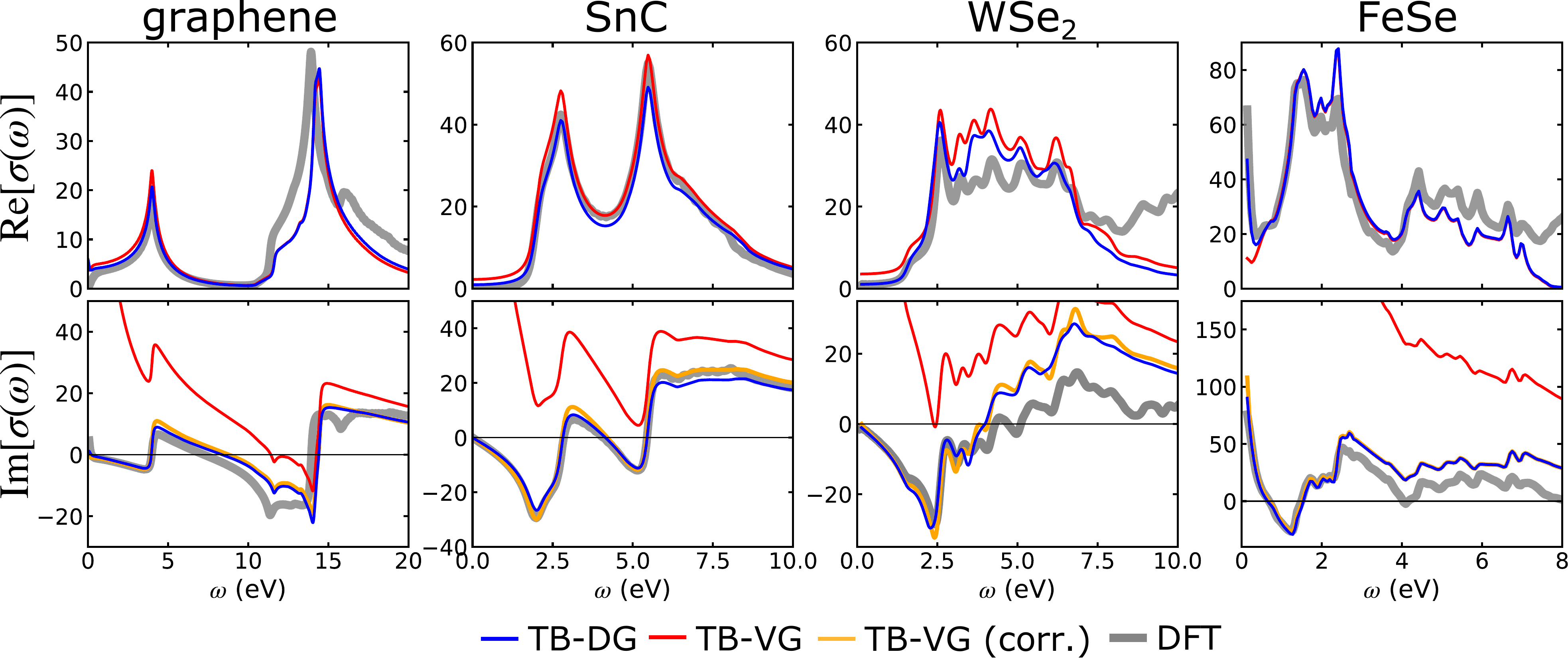}
\caption{Longitudinal optical conductivity of the considered 2D systems obtained from time propagation of the TB Hamiltonian in dipole (TB-DG) and velocity gauge (TB-VG), respectively. We propagated until a maximum time of $T_\mathrm{max}=8000$~a.u. and used $256\times 256$ sampling of the Brillouin zone, ensuring convergences. For the calculation of the conductivity with DFT we used a $200\times 200$ grid of Brillouin zone. We checked the convergence of the spectra in the considered frequency range with respect to the number of included bands. \label{fig:cond_wan}}
\end{figure*}\

\subsection{Optical conductivity\label{subsec:optcond}}

In the linear response regime, the current flowing through the system upon irradiation with light is fully determined by the optical conductivity $\sigma_{\mu\nu}(\omega)$. Solving the equation of motion for the SPDM in the velocity ($\gvec{\rho}(\vec{k},t)$) and the dipole gauge ($\widetilde{\gvec{\rho}}(\vec{k,t})$) and calculating the corresponding current \eqref{eq:total_current_velo} and \eqref{eq:curr_dip_wan} provide a direct route to computing the optical conductivity. To this end, we apply a short pulse of the form
\begin{align}
	\label{eq:pulse}
	\vec{E}(t) = \vec{e}\frac{F_0}{\sqrt{2\pi \tau^2}} e^{-\frac{t^2}{2 \tau^2}} \ ,
\end{align}
where $\vec{e}$ denotes the polarization vector. In the limit $\tau\rightarrow 0$, the pulse~\eqref{eq:pulse} becomes $\vec{E}(t)=\vec{e} F_0 \delta(t)$, containing all frequencies. Exploiting the linear relation between $\sigma_{\mu\nu}(\omega)$ and the electric field~\eqref{eq:pulse} upon $F_0 \rightarrow 0$, the optical conductivity is obtained by
\begin{align}
	\sigma_{\mu \nu}(\omega) = \frac{1}{F_0} e^{\omega^2\tau^2/2}\int^{\infty}_0\!dt\, e^{i \omega t} e^{-\eta t} J_{\mu}(t) \ ,
\end{align}
where $J_\mu(t)$ is the current in direction $\mu$ induced by choosing the polarization $\vec{e}$ along direction $\nu$. Here we focus on the longitudinal conductivity 
\begin{align}
	\label{eq:optcond_lon}
	\sigma(\omega) = \sigma_{xx}(\omega)  + \sigma_{yy}(\omega)  \ .
\end{align}
The damping factor $\eta$ is introduced for convergence, giving rise to Lorentzian broadening of the resulting spectra.

As a benchmark reference we calculated the optical conductivity using the program {\tt epsilon.x} from the \textsc{Quantum Espresso} package, which calculates the velocity matrix elements~\eqref{eq:velo_elemk} directly from the plane-wave representation of the Bloch wave-functions. Note that this procedure omits pseudopotential contributions to the velocity operator (which are neglected throughout this paper). We used Lorentzian smearing for both interband and intraband transitions, matching the parameter $\eta$ from the TB calculations. This procedure yields the dielectric function $\epsilon_{\mu\nu}(\omega)$, from which we calculate the longitudinal conductivity via $\sigma(\omega) = -i \omega (\epsilon_{xx}(\omega) + \epsilon_{yy}(\omega) - 2)$. This procedure amounts to the independent-particle approximation to the response properties.

\subsubsection{Conductivity within the velocity gauge vs. dipole gauge}

We solved the equation of motion for the SPDM with the Hamiltonian~\eqref{eq:sp_ham_velo} and computed the current according to Eq.~\eqref{eq:total_current_velo}. The velocity matrix elements were computed from the Wannier input via Eq.~\eqref{eq:velo_elemk_2}. We refer to the thus obtained results in the velocity gauge as TB-VG. Analogously, we have propagated the SPDM with the Hamiltonian~\eqref{eq:spham_trans} and computed the current according to Eq.~\eqref{eq:curr_dip_wan}. This defines the TB dipole gauge (TB-DG).

In Fig.~\ref{fig:cond_wan} we compare the optical conductivity from the TB models to the first-principle spectra. In general, the agreement for low-energy features (for which the TB models have been optimized) is very good for the real art. The major differences between the TB-DG model and TB-VG model is the unphysical behavior of $\mathrm{Im}[\sigma(\omega)]$ for $\omega \rightarrow 0$ in the velocity gauge. The TB-VG model displays a $\omega^{-1}$ behavior (albeit less pronounced for WSe$_2$). This artifact can be traced back to the violation of the sum rule~\eqref{eq:para_resp_free_fsum}. Larger deviations from $f=n$ lead to larger deviations from the reference conductivity. To check this behavior, we have evaluated $f$ according to Eq.~\eqref{eq:para_resp_free_fsum} (see Tab.~\ref{tab:fsum}). Including more empty bands leads to an improvement in the sum rule and thus in the behavior at small frequencies.
Inspecting the band structure (Fig.~\ref{fig:bands_wan}), we see that including even more bands above the Fermi energy into the TB models in not feasible, as higher excited states can hardly be described by localized WFs. In particular, for energies larger than the continuum threshold, the Bloch states are entirely delocalized. Achieving convergence of $\mathrm{Im}[\sigma(\omega)]$ within the TB-VG model is out of reach.

% \textcolor{red}{
% For instance, $f = 1.1$ for WSe$_2$, which is very close to the number of electrons (per spin) in the respective TB model, $n=1$, leads to a small downturn of $\mathrm{Im}[\sigma(\omega)]$ for $\omega\rightarrow 0$, while the deviation of $f$ and $n$ is larger for the other sytems.}

\begin{table}[b]
\caption{Overview of the TB models, number of electrons per unit cell (per spin) $n$, and the sum $f$ calculated from Eq.~\eqref{eq:para_resp_free_fsum}. For FeSe, the values have been obtained from the weight of the $\omega^{-1}$ term. \label{tab:fsum}}
\centering
\begin{ruledtabular} 
\begin{tabular}{llll}
system & \# of bands & $n$ & $f$ \\
\hline
graphene & 5 & 4 & 0.64 \\
         & 9 & 4 & 2.06 \\
SnC      & 6 & 4 & 1.94 \\
         & 8 & 4 & 2.45 \\
WSe$_2$  & 5 & 1 & 1.11 \\
		 & 11 & 7 & 4.18 \\
FeSe 	 & 10 & 6 & 2.10$^*$ \\
         & 16 & 12 & 3.06$^*$ \\
\end{tabular}
\end{ruledtabular}
\end{table}

However, imposing the correct $\omega^{-1}$ behavior is possible. Note that the divergence at small frequencies is solely due to the diamagnetic current, which is not canceled by the paramagnetic current. The cancellation (and thus the sum rule~\eqref{eq:para_resp_free_fsum}) can be enforced by replacing 
\begin{align}
	\label{eq:dia_corr}
	\vec{J}^\mathrm{dia}(t) = -q n \vec{A}(t) \rightarrow \vec{J}^\mathrm{dia,c}(t) =  -q f \vec{A}(t) \ .
\end{align}
Calculating the thus corrected current in the velocity gauge $\vec{J}^\mathrm{VG,c}(t) = \vec{J}^\mathrm{p}(t) + \vec{J}^\mathrm{dia,c}(t)$ defines the corrected TB-VG model.  The corrected model leads to excellent agreement between the dipole and the velocity gauge and cures the spurious $\omega^{-1}$ behavior $\mathrm{Im}[\sigma(\omega)]$  in all cases. There is no influence on $\mathrm{Re}[\sigma(\omega)]$. While the sum rule~\eqref{eq:para_resp_free_fsum} applies to insulators, incomplete cancellation of the paramagnetic and the diamagnetic current will also affect $\mathrm{Im}[\sigma(\omega)]$ for metallic systems like FeSe. In this case, we determine the $\omega^{-1}$ weight by $\omega \mathrm{Im}[\sigma(\omega)] \rightarrow 0$ and determine $f$ accordingly.

\subsubsection{Tight-binding vs. first-principle conductivity}

Inspecting the real part of the conductivity for graphene, we notice excellent agreement of the TB results with the first-principle spectrum, especially for energies $\omega < 10$~eV. For larger energies, the differences in the band dispersions gives rise to shifted spectra. Note that $\mathrm{Re}[\sigma(\omega)]\rightarrow 0$ is the exact behavior~\cite{stauber_optical_2008}, although the transition from almost constant $\mathrm{Re}[\sigma(\omega)]$ to 0 as $\omega\rightarrow 0$ is very abrupt and easily masked by smearing. Capturing this subtle feature is especially hard when calculating the conductivity from the time evolution of the current, as zero-frequency behavior is only accessible in the limit $t\rightarrow\infty$. We note that TB-VG and TB-DG are in excellent agreement. 

For SnC, all methods agree very well for the entire considered frequency range. Note the system is an insulator (at low temperature), so $\mathrm{Re}[\sigma(\omega)]\rightarrow 0$ for $\omega\rightarrow 0$. This is not exactly reproduced by the TB models (TB-VG is slightly worse); however, this can be cured by systematically increasing $T_\mathrm{max}$ and reducing the broadening $\eta$. Note that this procedure also requires finer sampling of the Brillouin zone. Besides the real part, also the imaginary part with the TB-DG and corrected TB-VG are in excellent agreement with the first-principle calculation. 

For WSe$_2$, the main absorption peak is well captured by the TB models (the TB-DG in particular). Similar to SnC, $\mathrm{Re}[\sigma(\omega)]$ does not tend to zero exactly for $\omega\rightarrow 0$. This behavior is consistently more pronounced with the TB-VG model. There are larger deviations of the imaginary part for $\omega > 3$~eV, which is to be expected from differences in peak structure of the real part due to the Kramers-Kronig relation. 

In contrast to the previous examples, FeSe is a metal. Due to the broadening used for all methods (which acts as a generic damping mechanism), the Drude peak is smeared out, giving rise to finite $\mathrm{Re}[\sigma(\omega)]$ for $\omega \rightarrow 0$. Again, the behavior for very small frequencies is well captured by the TB-DG model, while the TB-VG has difficulties for the chosen $\eta$ and the propagation time $T_\mathrm{max}$. Apart from the range $\omega\approx 0$, both TB models produce almost identical results, especially for the imaginary part (using the corrected TB-VG model). 

We have also computed $\sigma(\omega)$ for the reduced TB models (dashed lines in Fig.~\ref{fig:bands_wan}), presented in Appendix~\ref{app:reduced}. Comparing full and reduced models one finds that the artificial finite value of $\mathrm{Re}[\sigma(\omega)]$ for $\omega \rightarrow 0$ for insulating systems (within the TB-VG model) is less pronounced if $f \approx n$. Especially for WSe$_2$ ($f = 1.1$ within the reduced model, see Tab.~\ref{tab:fsum}), TB-DG and TB-VG model are almost identical. 

\subsection{Berry curvature \label{subsec:berry}}

\begin{figure}[t]
\centering
\includegraphics[width=\columnwidth]{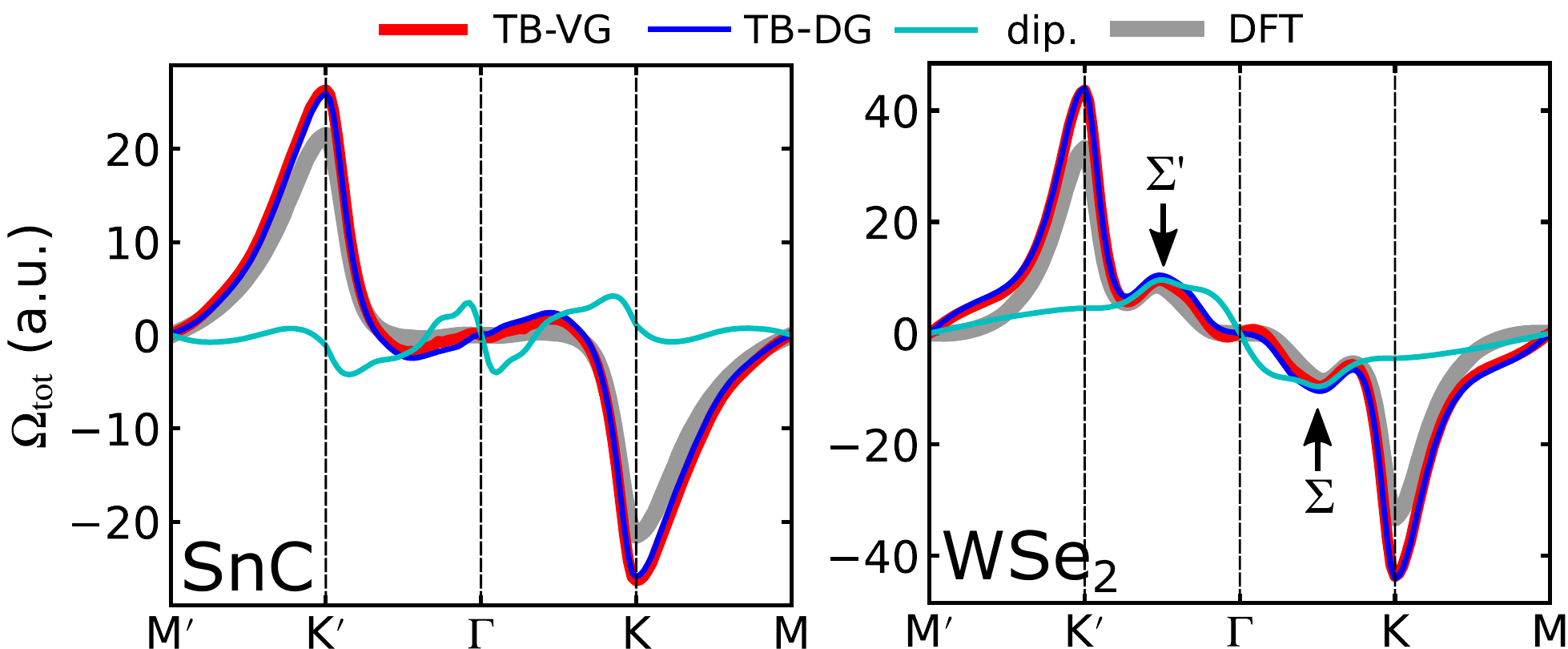}
\caption{Total Berry curvature of SnC (left) and WSe$_2$ (right panel) along a characteristic path in the Brillouin zone with the TB-DG model (Eq.~\eqref{eq:berry_disp}--\eqref{eq:berry_dip}), TB-VG model (Eq.~\eqref{eq:kubo_berry} and Eq.~\eqref{eq:velo_elemk_2}), and directly from the Bloch states (DFT). The turquoise line corresponds to the dipole contribution~\eqref{eq:berry_dip}. \label{fig:berry}}
\end{figure}

The described way of obtaining the optical conductivity can, of course, also be applied to the transverse response. In general, the Hall conductance $\sigma_{H} = \sigma_{xy}(\omega=0)$ of insulating systems contains information about their topological state due its close connection to the Berry curvature~\cite{yao_first_2004}:
\begin{align}
	\label{eq:sigma_hall}
	\sigma_H = \frac{e^2}{\hbar} \int_{\mathrm{BZ}} \frac{d \vec{k}}{(2\pi)^2}f_\alpha(\vec{k}) \Omega_\alpha(\vec{k}) \ .
\end{align}
Here, $\Omega_\alpha(\vec{k})$ is the Berry curvature of band $\alpha$. Exploiting Eq.~\eqref{eq:sigma_hall} and working out the paramagnetic linear response function~\eqref{eq:para_resp} explicitly (in the velocity gauge) yields the Kubo formula~\footnote{We restrict ourselves to the nondegenerate case here. The corresponding non-abelian expressions can be derived analogously~\cite{gradhand_first-principle_2012}.} for the Berry curvature~\cite{thouless_quantized_1982} in terms of the velocity matrix elements:
\begin{align}
	\label{eq:kubo_berry}
	\Omega_\alpha(\vec{k})= - 2 \mathrm{Im} \sum_{\alpha^\prime \ne \alpha} \frac{v^x_{\alpha \alpha^\prime}(\vec{k})v^y_{\alpha^\prime \alpha}(\vec{k})}{(\en_\alpha(\vec{k}) - \en_{\alpha^\prime}(\vec{k}))^2}\ .
\end{align}
In practice, the velocity matrix elements are usually computed from the Wannier representation and Eq.~\eqref{eq:velo_elemk_2}. However, the formulation of the real-time dynamics in terms of the dipole gauge provides an alternative route. To this end we evaluate the current~\eqref{eq:curr_dip_wan} in linear response. Inserting into the  current-current response function and evaluating the corresponding conductivity, one obtains two distinct contributions: $\Omega_\alpha(\vec{k}) = \Omega^\mathrm{disp}_\alpha(\vec{k}) + \Omega^\mathrm{dip}_\alpha(\vec{k})$. This is in direct analogy to the current contributions~\eqref{eq:curr_dip_wan_disp} and \eqref{eq:pol_wan}. 
The dispersion part reads
\begin{widetext}
\begin{align}
\label{eq:berry_disp}
 \Omega^\mathrm{disp}_\alpha(\vec{k}) = -2 \mathrm{Im} \sum_{\alpha^\prime \ne \alpha} \frac{\vec{C}^\dagger_\alpha(\vec{k}) \partial_{k_x} \widetilde{\vec{h}}(\vec{k}) \vec{C}_{\alpha^\prime}(\vec{k}) \vec{C}^\dagger_{\alpha^\prime}(\vec{k}) \partial_{k_y} \widetilde{\vec{h}}(\vec{k}) \vec{C}_{\alpha}(\vec{k})}{(\en_\alpha(\vec{k}) - \en_{\alpha^\prime}(\vec{k}))^2} \ , 
\end{align}
while for the dipole part one finds
\begin{align}
\label{eq:berry_dip}
 \Omega^\mathrm{dip}_\alpha(\vec{k}) = 2 \mathrm{Re} \sum_{\alpha^\prime \ne \alpha} \left(\frac{\vec{C}^\dagger_\alpha(\vec{k}) \partial_{k_x} \widetilde{\vec{h}}(\vec{k}) \vec{C}_{\alpha^\prime}(\vec{k}) }{\en_\alpha(\vec{k}) - \en_{\alpha^\prime}(\vec{k})} D^y_{\alpha \alpha^\prime}(\vec{k})- \frac{\vec{C}^\dagger_\alpha(\vec{k}) \partial_{k_y} \widetilde{\vec{h}}(\vec{k}) \vec{C}_{\alpha^\prime}(\vec{k}) }{\en_\alpha(\vec{k}) - \en_{\alpha^\prime}(\vec{k})} D^x_{\alpha \alpha^\prime}(\vec{k})\right) \ .
\end{align}
\end{widetext}
For brevity, we have introduced the vector notation $[\vec{C}_\alpha(\vec{k})]_m = C_{m\alpha}(\vec{k})$, while $D^{\mu}_{\alpha \alpha^\prime}(\vec{k}) = \sum_{mn} C^*_{m\alpha}(\vec{k})D^\mu_{m n}(\vec{k}) C_{n \alpha}(\vec{k})$ denotes the dipole matrix elements in the Bloch basis. The expressions~\eqref{eq:berry_disp} and \eqref{eq:berry_dip} are equivalent to Eq.~(71)--(72) from ref.~\cite{gradhand_first-principle_2012}. Assuming a complete basis of WFs one can also obtain Eq.~\eqref{eq:berry_disp}--\eqref{eq:berry_dip} from Eq.~\eqref{eq:kubo_berry} inserting Eq.~\eqref{eq:velo_elemk_2} and \eqref{eq:berryconect_2}. For an incomplete basis the equivalence is only guaranteed if $\sum_{n} D^\mu_{mn}(\vec{k})D^\nu_{n m^\prime}(\vec{k}) = \sum_{n} D^\nu_{mn}(\vec{k})D^\mu_{n m^\prime}(\vec{k}) $, i\,e. if the dipole operators with respect to orthogonal directions commute. 

\begin{figure*}[t]
\center
\includegraphics[width=\textwidth]{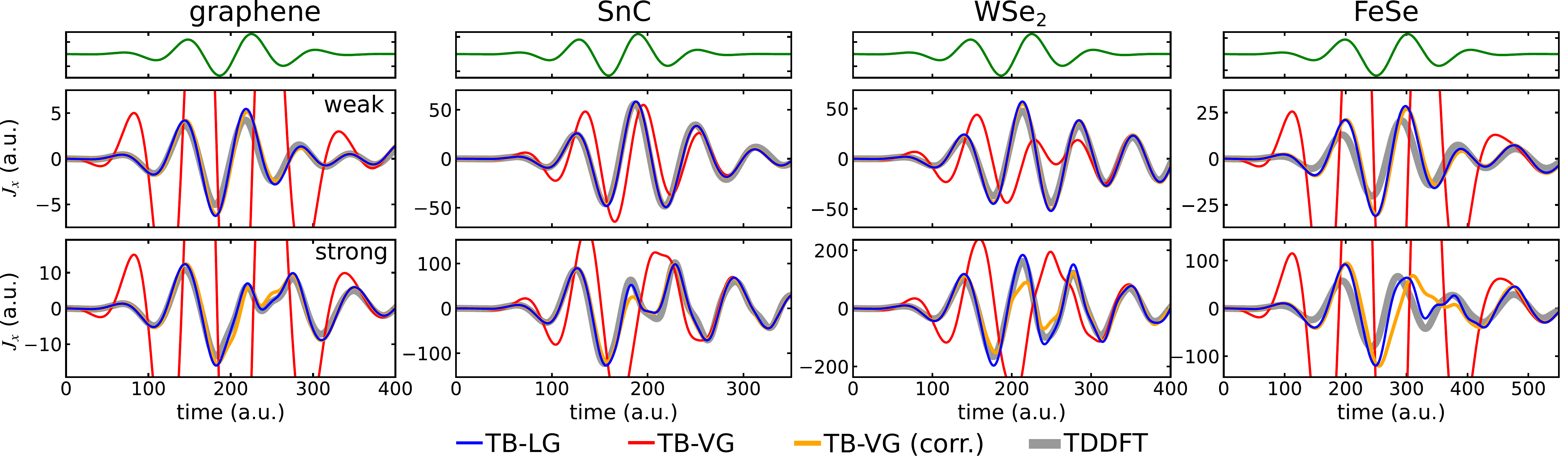}
\caption{Current induced by a few cycle pulse (electric field shown in top panels) in the case of weak (middle) and strong driving (bottom panels). Calculations were performed with a $48\times 48$ sampling of the Brillouin zone for all cases. For better readability, $J_x$ has been multiplied by the factor $10^3$. \label{fig:pulse_wan}}
\end{figure*}

We have calculated the Berry curvature for the two systems that break inversion symmetry -- SnC and WSe$_2$ -- (i) from Eq.~\eqref{eq:kubo_berry} inserting the velocity matrix elements~\eqref{eq:velo_elemk_2} from the respective TB model, (ii) from Eq.~\eqref{eq:berry_disp}--\eqref{eq:berry_dip}, and (iii) from Eq.~\eqref{eq:kubo_berry} based on velocity matrix elements calculated from the Bloch states directly. We have used the \textsc{Octopus} code to compute the matrix elements from the real-space representation of the $\psi_{\vec{k}\alpha}(\vec{r})$ and the momentum operator $\hat{\vec{p}}=-i \nabla_{\vec{r}}$. Converging the obtained Berry curvature with respect to the number of bands thus serves as a benchmark.

In Fig.~\ref{fig:berry} we compare the different models for calculating the total Berry curvature $\Omega_\mathrm{tot}(\vec{k}) = \sum_{\alpha} f_{\alpha}(\vec{k})\Omega_{\alpha}(\vec{k})$. For SnC, $\Omega_\mathrm{tot}(\vec{k})$ is almost identical within the TB-DG and TB-VG model. Both models agree qualitatively with the DFT calculation, albeit the magnitude of the Berry curvature is slightly overestimated specifically in the vicinity of the K and K$^\prime$ point. This is explained by all (dipole-allowed) bands, especially higher conduction bands that are missing in the TB models, contributing to the Berry curvature. 
The picture is similar for WSe$_2$. Interestingly, the peak of the Berry curvature between K (K$^\prime$) and $\Gamma$ (called $\Sigma$ ($\Sigma^\prime$) valley) is well reproduced by both gauges. We also show the dipole contribution~\eqref{eq:berry_dip}. While for both materials the dispersion part~\eqref{eq:berry_disp} dominates, for WSe$_2$ the dipole part is the predominant contribution close to $\Sigma^{(\prime)}$ valley. Note that this feature would be missed by the usual TB models~\cite{fang_textitab-initio_2015} that are constructed without the dipole matrix elements.

\subsection{Nonlinear dynamics \label{sec:nonlin} }

We proceed to investigating the nonlinear response. To this end we simulated the dynamics upon a short laser pulse, defined by
\begin{align}
	\label{eq:pulse}
	\vec{A}(t) = \vec{e}_x A_0 \exp\left(-a \left(\frac{t-t_0}{\tau}\right)^2\right)\cos[\omega_0(t-t_0)] \ .
\end{align}
Here, $\vec{e}_x$ denotes the unit vector in $x$ direction. Choosing the parameters $a =4.6$ and $\tau=2\pi n_c / \omega_0$, the vector potential~\eqref{eq:pulse} represents an $n_c$-cycle pulse. We choose $n_c=2$ and determine $\omega_0$ to drive typical excitations within the band manifold spanned by the TB models. For the pulse strengh $A_0$ we consider two scenarios: (i) weak driving (but beyond linear response), and (ii) strong excitation. We have chosen $A_0$ to obtain representative examples of the dynamics. Tab.~\ref{tab:param} lists the pulse parameters for all systems considered.

\begin{table}[b]
\caption{Pulse parameters defining the pulse~\eqref{eq:pulse} used for the simulations. $\omega_0$ is given in units of eV, while atomic units are used for $A_0$. \label{tab:param}}
\centering
\begin{ruledtabular} 
\begin{tabular}{llll}
system & $\omega_0$ & $A_0$ (weak) & $A_0$ (strong) \\
\hline
graphene & 2.0 & 0.025 & 0.075 \\
SnC      & 2.5 & 0.035 & 0.125 \\
WSe$_2$  & 2.0 & 0.025 & 0.125 \\
FeSe 	 & 1.5 & 0.025 & 0.1125 \\
\end{tabular}
\end{ruledtabular}
\end{table}

To obtain an accurate benchmark, we have simulated the dynamics with the TDDFT code \textsc{Octopus}. The Kohn-Sham potential was frozen as the ground-state potential, so that all calculations are performed on equal footing (independent particle approximation). The current was calculated (ignoring pseudopotential corrections) from Eq.~\eqref{eq:current_rsp}.

\begin{figure}[b]
\center
\includegraphics[width=\columnwidth]{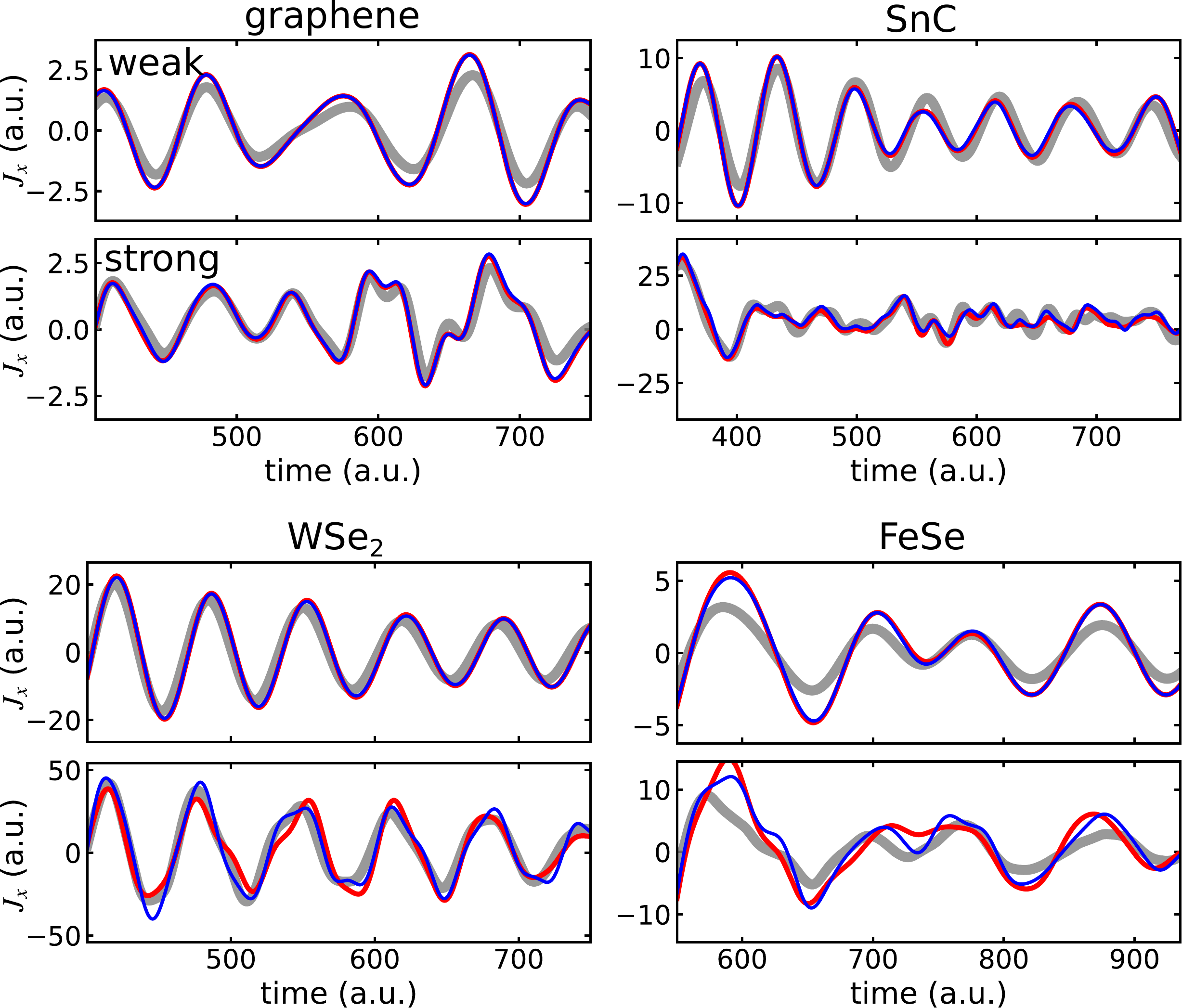}
\caption{Current induced by a few cycle pulse as in Fig.~\ref{fig:pulse_wan} for longer times $t\gg \tau$ where $\vec{A}(t) \approx 0$. The color coding is consistent with Fig.~\ref{fig:pulse_wan}. \label{fig:pulseafter_wan}}
\end{figure}

In Fig.~\ref{fig:pulse_wan} we present the induced current $J_x$ along with the shape of the pulse~\eqref{eq:pulse} for all systems. Comparing the TB-VG model to the TDDFT results one finds pronounced differences, especially for graphene and FeSe. The oscillations of $J_x$ seem out of phase. This behavior is due to the incomplete cancellation of paramagnetic and diamagnetic current, similar to the linear response case. It is less severe for SnC and WSe$_2$, where the sum rule $f=n$ (cf. Eq.~\eqref{eq:para_resp_fsum}) is violated to a lesser extent. Following the same procedure as in Section~\ref{subsec:optcond}, we replaced $n \rightarrow f$ when calculating the diamagnetic current~\eqref{eq:dia_corr}. The thus obtained corrected TB-VG model yields almost identical results as the TB-DG model in the regime of weak driving (albeit it is beyond linear response). Even for strong driving, the corrected TB-VG model is in good agreement with the TB-DG model, although deviations become apparent when the field peaks. The TB-DG model reproduces the TDDFT current better than the TB-VG model. Even for strong excitations, the TB-DG model is in remarkably good agreement.

For all systems, both TB models yield a very good approximation to the current $J_x(t)$ obtained from TDDFT. The agreement is particularly good for SnC and WSe$_2$. For FeSe the magnitude of the the current is slightly overestimated~\footnote{Inspecting the velocity matrix elements obtained from the Bloch states directly and from the TB models shows some qualitative discrepencies.}

We have also performed analogous simulations for the reduced TB models. As expected, for weak excitation the results are almost identical to Fig.~\ref{fig:pulse_wan}, while deviations are more pronounced for stronger driving. 
This is particularly pronounced for WSe$_2$. In this case, the lack of the Se $p$ bands (see Fig.~\ref{fig:bands_wan}) limit the nonresonant optical transitions, and the $p$-$d$ hybdrization of the $d_{xz}$ and $d_{yz}$ orbitals is missing. For graphene and SnC, excluding higher bands has only minor effect, as the additional bands are strongly off-resonant. For FeSe, excluding the lower-lying $p$ bands has almost no noticeable effect for the same reason.

There is still current flowing in the systems after the pulse is essentially zero ($t \gg \tau$), which is mostly due to the induced oscillations of the dipole moments. Fig.~\ref{fig:pulseafter_wan} shows the current corresponding to Fig.~\ref{fig:pulse_wan} in this field-free regime. As $\vec{A}(t)$ is vanishingly small, there is no diamagnetic contribution spoiling the TB-VG model. Both the TB-DG and TB-VG model are in very good agreement with the TDDFT calculation, albeit the TB-DG model seems to have a slight edge over the TB-VG model. 

\section{Conclusions}

We have studied light-induced dynamics in 2D systems in the linear response regime and beyond, focusing on the different ways of introducing light-matter interaction. While all gauges of light-matter coupling are connected by a unitary transformation of the Hamiltonian and are thus equivalent, from a practical point of view it pertinent to assess the accuracy of each gauge. This is particularly important when working with TB models to capture low-energy excitations, which inherently breaks the completeness relation.
We have introduced the standard velocity gauge from the minimal coupling scheme and presented the transformation to the dipole gauge, which yields a multi-band extension of the Peierl's substitution. To systematically investigate the performance of dipole or velocity gauge in a reduced basis, we have constructed first-principle TB models including dipole transition matrix elements in Wannier basis. As an accurate reference method, we performed TDDFT simulations with a converged plane-wave (for linear response) or real-space (for nonlinear dynamics) basis.

Linear response properties -- we focused on the optical conductivity -- are well captured by the TB models in their corresponding energy range. The TB-VG model, however, shows spurious $\omega^{-1}$ behavior for the imaginary part of the conductivity, which can be traced back to a violation of a sum rule of the paramagnetic response function. In contrast, the TB-DG model captures the correct low-frequency behavior by construction. Correcting the TB-VG model by hand is possible by enforcing the sum rule. Instead of correcting the TB-VG model, convergence of the low-energy behavior can be achieved by (i) systematically increasing the number of conduction bands, and (ii) excluding lower-lying valence bands that are not participating in the dynamics. This strategy is exemplified by WSe$_2$. However, the delocalized nature of higher conduction bands renders (i) impractical. Thus, enforcing the paramagnetic sum rule is a more efficient way of systematically improving the imaginary part of the conductivity.
This procedure will also be important when investigating light-matter interaction beyond the dipole approximation (like Raman or X-ray scattering), where the diamagnetic term is responsible for the excitations. 

TB models are also a convenient way to calculate topological properties like the Berry curvature. With an accurate Wannier representation to the Bloch states, the Berry curvature is almost identical within the TB-VG and TB-DG model. The dipole-gauge formulation furthermore allows to disentangle orbital hybridization and dipole couplings. The latter contribution, which is often ignored in the TB framework, can be important as demonstrated for WSe$_2$.

Nonlinear excitations can also be captured accurately within the TB models. Similar to the linear response case, the lack of cancellation of paramagnetic and diamagnetic current within the TB-VG model gives rise to a strongly overestimated total current. Remarkably, enforcing the cancellation on the linear-response level cures these deficiencies even for strong excitations. The TB-DG model provides a more accurate description, especially for strong pulses.

In summary, both the TB-VG and the TB-DG model provide an excellent description of light-induced dynamics (as long as the relevant bands are included), along with all the advantages of TB model: simplicity, low computational cost, straightforward interpolation to any momentum grid, and the possibility to include many-body effects with quantum kinetic methods.

\section*{Acknowledgments}
We acknowledge insightful discussions with Denis Gole\v{z}, Brian Moritz and C. Das Pemmaraju. 
We also thank the Stanford Research Computing Center for providing computational resources. Data used in this manuscript is stored on Stanford's Sherlock computing cluster. Supported by the U.S. Department of Energy (DOE), Office of Basic Energy Sciences, Division of Materials Sciences and Engineering, under contract DE-AC02-76SF00515.
M.~S. thanks the Alexander von Humboldt
Foundation for its support with a Feodor Lynen scholarship.
Y.~M. acknowledges the support by a Grant-in-Aid for Scientific Research from JSPS, KAKENHI Grant Nos. JP19K23425, JP20K14412, JP20H05265 and JST CREST Grant No. JPMJCR1901.

\appendix

\section{Transformation to the dipole gauge\label{app:trans_len}}

For completeness we present the detailed derivation of the dipole gauge in this appendix. We start from the minimal-coupling Hamiltonian in Wannier representation, defined by
\begin{align}
\label{eq:ham_sp_wann}
	T_{m\vec{R}n\vec{R}^\prime}(t) = \big\langle m \vec{R} \big| \frac{1}{2}(\hat{\vec{p}}-q \vec{A}(t))^2 + \hat{v} \big| n \vec{R}^\prime\big\rangle  \ .
\end{align}
For finite systems the unitary transformation is constructed from the generator $\hat{S}(t) = -i q \vec{A}(t)\cdot \vec{r}$, i.\,e. $\hat{U}(t) = e^{\hat{S}(t)}$. Expressing the dipole operator in the Wannier basis, the generalization of this generator to periodic system is defined by
\begin{align}
\label{eq:dipole_gen}
S_{m\vec{R}n\vec{R}^\prime} = -\iu q \vec{A}(t)\cdot \sum_{\vec{R}\vec{R}^\prime} \sum_{m n} 
\vec{D}_{m\vec{R}n\vec{R}^\prime} \ .
\end{align}
Collecting orbital and site indices in compact matrix notation, the generator~\eqref{eq:dipole_gen} defines the unitary transformation $\vec{U}(t) = e^{\vec{S}(t)}$ by its matrix elements (cf. Eq.~\eqref{eq:unitary_len}).
The generator must obey $\vec{S}(t) = -\vec{S}^\dagger(t)$ to define a unitary transformation, which is fulfilled if $\vec{D}_{m\vec{R}m^\prime\vec{R}^\prime} = \vec{D}^*_{m^\prime\vec{R}^\prime m\vec{R}}$. In analogy to Eq.~\eqref{eq:len_finite}, the dipole-gauge Hamiltonian in Wannier representation is obtained by transforming Eq.~\eqref{eq:ham_sp_wann}:
\begin{align}
	\label{eq:ham_wan_len}
	\widetilde{h}_{m\vec{R}m^\prime \vec{R}^\prime}(t) = [\vec{U}(t) (\vec{T}(t)+ \iu \partial_t \vec{S}(t) ) \vec{U}^\dagger(t)]_{m\vec{R}m^\prime\vec{R}^\prime} \ .
\end{align}
The first term gives rise to the Peierl's phase factor
\begin{widetext}
\begin{align*}
\widetilde{T}_{m\vec{R} m^\prime\vec{R}^\prime}(t) &= 
    [\vec{U}(t) \vec{T}(t) \vec{U}^\dagger(t)]_{m\vec{R}m^\prime\vec{R}^\prime} = \sum_{\vec{R}_1\vec{R}_2}\sum_{n_1 n_2} \langle m \vec{R}|e^{-i q \vec{A}(t)\cdot(\vec{r}-\vec{R})}| n_1 \vec{R}_1 \rangle 
    T_{n_1\vec{R}_1 n_2\vec{R}_2}(t) \langle n_2 \vec{R}_2 | e^{i q \vec{A}(t)\cdot(\vec{r}-\vec{R}^\prime)}
    | m^\prime \vec{R}^\prime \rangle 
    \\
	 &= e^{\iu q \vec{A}(t)\cdot(\vec{R}-\vec{R}^\prime)} \langle m \vec{R} | e^{-\iu q \vec{A}(t)\cdot\vec{r}} \left(\frac12 \left(\hat{\vec{p}}-q \vec{A}(t)\right)^2 + \hat{v}\right)  e^{\iu q \vec{A}(t)\cdot\vec{r}} | m^\prime \vec{R}^\prime \rangle = e^{\iu q \vec{A}(t)\cdot(\vec{R}-\vec{R}^\prime)} T_{m\vec{R}m^\prime\vec{R}^\prime} \ ,
\end{align*}
\end{widetext}
while the second term arises due to the time-dependence of the generator~\eqref{eq:dipole_gen}. Following similar steps as above, we can show
\begin{align*}
[\vec{U}(t) \iu \partial_t \vec{S}(t) \vec{U}^\dagger(t)]_{m\vec{R} m^\prime\vec{R}^\prime} = -q \vec{E}(t)\cdot e^{\iu q \vec{A}(t)\cdot(\vec{R}-\vec{R}^\prime)}  \vec{D}_{m\vec{R}m^\prime\vec{R}^\prime}\ .
\end{align*}
Hence, the unitary transformation~\eqref{eq:ham_wan_len} gives rise to the Wannier Hamiltonian~\eqref{eq:ham_sp_dipole_new}.

It is straightforward to show that the SPDM obeys the transformed equation of motion 
\begin{align}
	\label{eq:eom_spdm_len}
	\frac{d}{d t} \widetilde{\gvec{\rho}}(\vec{k},t) = -i \left[ \widetilde{\vec{h}}(\vec{k},t), \widetilde{\gvec{\rho}}(\vec{k},t)  \right]  \ .
\end{align}
The dipole-gauge SPDM $\widetilde{\gvec{\rho}}(\vec{k},t)$ transforms according to $\widetilde{\rho}_{m \vec{R} m^\prime \vec{R}^\prime}(t) = [\vec{U}(t) \gvec{\rho}(t) \vec{U}^\dagger(t)]_{m \vec{R} m^\prime \vec{R}^\prime}$. Here $[\gvec{\rho}(t)]_{m \vec{R} m^\prime \vec{R}^\prime}$ denotes the velocity-gauge SPDM in Wannier representation.

\subsection{Total current in the dipole gauge\label{subsec:len_curr}}

\begin{figure*}[t]
\center
\includegraphics[width=\textwidth]{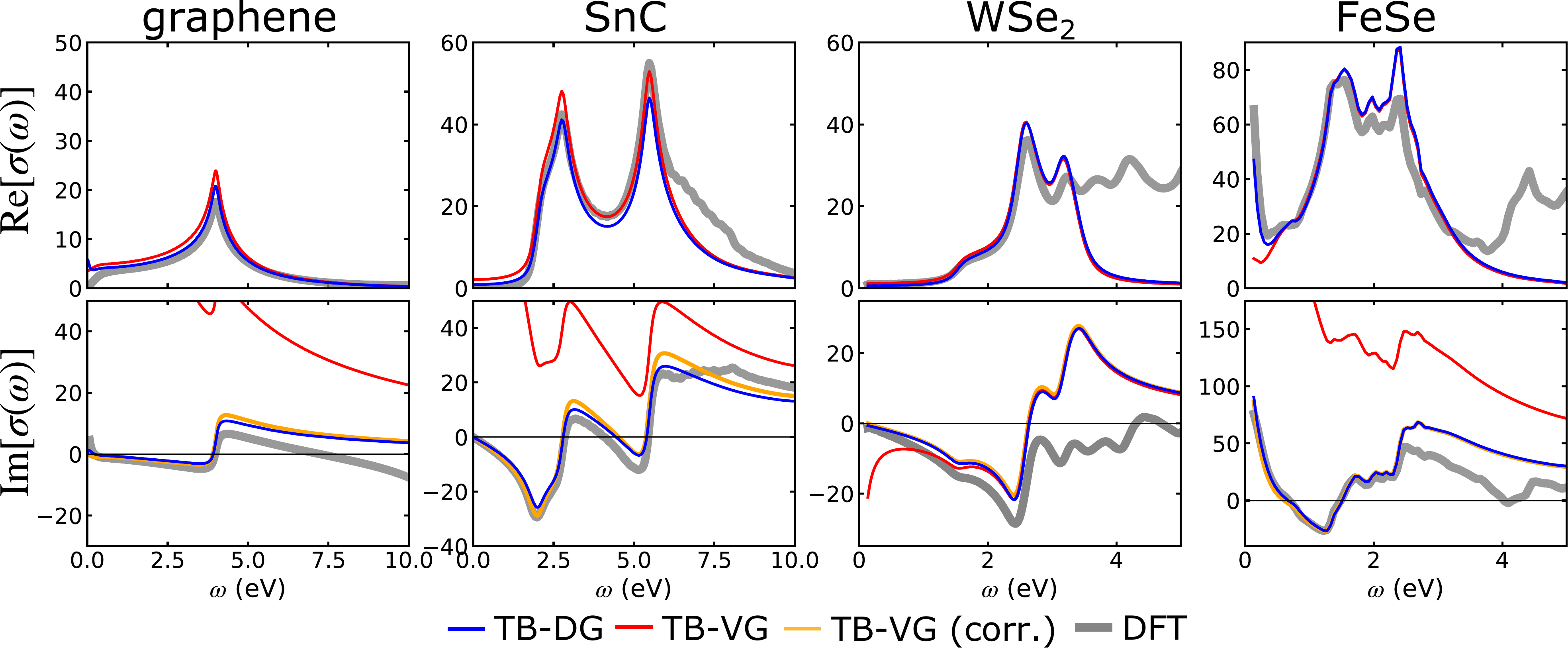}
\caption{Optical conductivity as in Fig.~\ref{fig:cond_wan} (using the same parameters), but for the reduced TB models.
 \label{fig:cond_wan_red}}
\end{figure*}

To derive the expression for the current in the dipole gauge, we start from the minimal coupling formulation~\eqref{eq:total_current_velo} and require gauge invariance. Switching to the Wannier basis, the expectation value of total current~\eqref{eq:total_current_velo} reads
\begin{align*}
	\vec{J}^\mathrm{VG}(t) = \frac{q}{N} \sum_{\vec{R} \vec{R}^\prime} \sum_{m m^\prime} \langle m \vec{R} | \hat{\vec{p}} - q \vec{A}(t)| m^\prime \vec{R}^\prime \rangle \rho_{m^\prime \vec{R}^\prime m \vec{R}}(t) \ ,
\end{align*}
where $\rho_{m^\prime \vec{R}^\prime m \vec{R}}(t) = \sum_{\vec{k}} \rho_{mm^\prime}(\vec{k},t) e^{-i \vec{k}\cdot (\vec{R} - \vec{R}^\prime)}$ denotes the SPDM in Wannier basis. Exploiting the cyclic invariance of the trace, we insert the unitary transformation~\eqref{eq:unitary_len} to transform the momentum matrix elements and SPDM to the dipole gauge. One finds
\begin{align}
	\label{eq:curr_mom_wan}
	\vec{J}^\mathrm{LG}(t) = \frac{q}{N} \sum_{\vec{R} \vec{R}^\prime} \sum_{m m^\prime} e^{-i q \vec{A}(t)\cdot (\vec{R}-\vec{R}^\prime)} \langle m \vec{R} | \hat{\vec{p}}
	| m^\prime \vec{R}^\prime \rangle \widetilde{\rho}_{m^\prime \vec{R}^\prime m \vec{R}}(t) \ .
\end{align}
If the matrix elements of the momentum operator are available in the Wannier basis, Eq.~\eqref{eq:curr_mom_wan} provides a direct way of obtaining the (gauge-invariant) total current. However, it is typically more convenient to calculate dipole matrix elements instead. Note that this also how the Berry connection~\eqref{eq:berryconect_2} is computed based on WFs~\cite{yates_spectral_2007}. Therefore, we replace the momentum operator by $\hat{\vec{p}} = -i [\vec{r}, \hat{h}(t)]$. 

Using the cell-centered dipole matrix elements~\eqref{eq:dipole_op}, one finds 
\begin{widetext}
\begin{align*}
	\langle m \vec{R} | \hat{\vec{p}}
	| m^\prime \vec{R}^\prime \rangle = -i (\vec{R}-\vec{R}^\prime) T_{m\vec{R}m^\prime \vec{R}^\prime} -i \sum_{\vec{R}_1,n_1} \Big( D_{m\vec{R}n_1\vec{R}_1} T_{n_1\vec{R}_1 m^\prime \vec{R}^\prime} - T_{m \vec{R} n_1 \vec{R}_1} D_{n_1\vec{R}_1 m^\prime \vec{R}^\prime} \Big) \ .
\end{align*}
\end{widetext}
The structure suggests two distinct terms which contribute to the current: $\vec{J}(t) = \vec{J}^{(1)}(t) + \vec{J}^{(2)}(t)$. Fourier transforming the first term and the SPDM to momentum space, the first contribution simplyfies to
\begin{align}
	\label{eq:curr_dip_1}
	\vec{J}^{(1)}(t) = \frac{q}{N}\sum_{\vec{k}} \sum_{m m^\prime} \nabla_{\vec{k}} T_{m m^\prime}(\vec{k}-q \vec{A}(t)) \widetilde{\rho}_{m^\prime m}(\vec{k},t) \ .
\end{align}
Similarly, the second contribution after switching to momentum space is given by
\begin{align}
	\vec{J}^{(2)}(t) = \frac{q}{N} \sum_{\vec{k}} \sum_{m m^\prime} \gvec{\mathcal{P}}_{m m^\prime}(\vec{k}-q\vec{A}(t)) \widetilde{\rho}_{m^\prime m}(\vec{k},t) \ ,
\end{align}
where
\begin{align}
	\label{eq:curr_dip_p}
	\gvec{\mathcal{P}}_{m m^\prime}(\vec{k}) = -i \sum_n \left(\vec{D}_{m n}(\vec{k})T_{n m^\prime}(\vec{k}) 
	- T_{m n}(\vec{k})\vec{D}_{n m^\prime}(\vec{k})  \right) \ .
\end{align}
We note that $T_{m m^\prime}(\vec{k}-q\vec{A}(t))$ can be replaced by $\widetilde{h}_{m m^\prime}(\vec{k},t)$ in the commutator~\eqref{eq:curr_dip_p}. Using the identity $\mathrm{Tr}([A,B] C) = \mathrm{Tr}(A [B,C])$ one thus obtains
\begin{align}
\label{eq:curr_dip_2}
	\vec{J}^{(2)}(t) = \frac{q}{N}\sum_{\vec{k}} \sum_{m m^\prime} D_{m m^\prime}(\vec{k}-q\vec{A}(t)) \frac{d}{dt}
	\widetilde{\rho}_{m^\prime m}(\vec{k},t) \ .
\end{align}

We can move the time derivative from the density matrix to the whole expression by compensating the derivative acting on $\vec{D}_{mm^\prime}(\vec{k}-q\vec{A}(t))$. One thus obtains $\vec{J}^\mathrm{LG}(t) = \vec{J}^\mathrm{disp}(t) + \vec{J}^\mathrm{dip}(t)$, where the two contributions are defined by Eq.~\eqref{eq:curr_dip_wan_disp} and Eq.~\eqref{eq:pol_wan} (by $\vec{J}^\mathrm{dip}(t) = \dot{\vec{P}}(t)$), respectively.

\subsection{Static limit\label{subsec:len_static}}

For a insulating system at zero temperature, the DC current response vanishes. This property is fulfilled by construction in the dipole gauge. We note that the displacement current does not contribute to the DC current, as $\vec{J}^\prime_2(\omega) = -i \omega \vec{P}(\omega) \rightarrow 0$ for $\omega \rightarrow 0$, as $\vec{P}(\omega\rightarrow 0)$ stays finite. For showing that the static contribution $\vec{J}^\prime_1(\omega\rightarrow 0)$ vanishes, it is convenient to switch to a band basis:
\begin{align}
	\label{eq:curr_dip_1_w}
	\vec{J}^\prime_1(\omega) = \frac{q}{N} \int^\infty_{-\infty}\! d t\, e^{i \omega t} \sum_{\vec{k}} \sum_{\alpha \in \mathrm{occ}} \langle \psi_{\vec{k}\alpha}(t)| \nabla_{\vec{k}} \widetilde{\vec{h}}(\vec{k},t) | \psi_{\vec{k}\alpha}(t)\rangle
\end{align}
Employing first-order time-dependent perturbation theory to the time-dependent Bloch states $| \psi_{\vec{k}\alpha}(t)\rangle$ assuming a quasi-static electric field one finds that only valence bands can appear in the expansion $| \psi_{\vec{k}\alpha}(t)\rangle=\sum_{\nu} C_{\alpha\nu}(\vec{k},t) |\psi_{\vec{k}\nu}\rangle$. Using this property and expanding Eq.~\eqref{eq:curr_dip_1_w} up to linear order in the external fields one finds $\vec{J}^\prime_1(\omega) \rightarrow 0$ for $\omega\rightarrow 0$, similar to the single-band Peierl's substitution.

\section{Conductivity within reduced tight-binding models\label{app:reduced}}

We have computed the optical conductivity $\sigma(\omega)$ for the reduced TB models (see dashed lines in Fig.~\ref{fig:bands_wan}) for all systems by the same procedure as for the full models (see Sec.~\ref{subsec:optcond}). The result is shown in Fig.~\ref{fig:cond_wan_red}.

% \bibliographystyle{apsrev4-1}
% \bibliography{mylib}{}


\end{document}